\documentclass[pss]{wiley2sp}
\usepackage{amsmath}
\usepackage{bm}              
\usepackage{w-greek}         

\begin{document}



\title{Quantum Criticality and Global Phase Diagram of Magnetic Heavy Fermions}

\titlerunning{Quantum criticality and global phase diagram}

\author{%
  Qimiao Si 
}
\authorrunning{Q.~Si}

\mail{e-mail
  \textsf{qmsi@rice.edu}}

\institute{%
Department of Physics \& Astronomy, Rice University, Houston, TX 77005, USA\\
}



\abstract{%
%
%
%
\abstcol{
Quantum criticality describes the collective fluctuations of matter
undergoing a second-order phase transition at zero temperature.
It is being discussed in a number of strongly correlated electron
systems. A prototype case occurs in the heavy fermion metals, 
in which antiferromagnetic quantum critical points have been 
explicitly observed. Here, I address two types of antiferromagnetic
quantum critical points. 
In addition to the standard description
based on the fluctuations of the antiferromagnetic order, 
a local quantum critical point 
is also 
considered. 
}
{
It contains inherently quantum modes 
that are associated with a critical breakdown of the Kondo effect.
Across such a quantum critical point, there is a sudden 
collapse of a large Fermi surface 
to a small one.
I also consider the proximate antiferromagnetic and paramagnetic
phases, and these considerations lead to a global phase diagram. 
Finally, I discuss the pertinent experiments on the 
antiferromagnetic heavy fermions, briefly 
address the case of 
ferromagnetic heavy fermions, and 
outline some directions for future studies.
}}

\maketitle   

\section{Introduction}
A quantum critical point (QCP) refers to a second-order phase transition
at zero temperature. The notion of quantum criticality is playing a central
role in a number of strongly correlated systems, but this was not anticipated
when it was first introduced.
Indeed, the initial work of Hertz \cite{Hertz} 
was rather modest. 
From the critical phenomenon perspective, 
Hertz formulated a direct extension of 
Wilson's then-newly-completed renormalization-group (RG)
theory of classical critical phenomena \cite{Wilson.74}.
The formulation retained the basic property of the latter:
the zero-temperature phases are still considered to be 
distinguished by an order parameter,
a coarse-grained macroscopic variable characterizing 
the breaking of a global symmetry of the Hamiltonian,
and the critical modes are the fluctuations of the 
order parameter. In this sense, it retains the 
Landau paradigm for phase transitions.
From a microscopic perspective, Hertz's discussion 
built on the historical work about paramagnons, the overdamped 
magnons occurring in a paramagnetic metal as it becomes more and more
ferromagnetic (for a review, see Ref.~\cite{LevinValls.83}).
In hindsight, the convergence of these two lines of theoretical
physics
seems to be rather
natural. For a Stoner ferromagnet, the fluctuations of the 
order parameter (magnetization) at its QCP is none other 
but the paramagnons. For a spin-density-wave (SDW)
antiferromagnet, such fluctuations of the order 
parameter (staggered magnetization) are correspondingly 
antiparamagnons. For completeness, it is interesting to note 
on the third line of activities predating Hertz's work.
A QCP was already contained in the microscopic solution
of an Ising chain in a transverse field \cite{Pfeuty.70},
and this solution was being reformulated in the Landau 
framework \cite{Young.75}.

The order-parameter fluctuations of a classical magnetic critical
point is specified by a $\phi^4$ field theory 
in d-spatial dimensions \cite{Wilson.74}.
Within the Hertz formulation, quantum mechanics introduces 
mixing between the statics and dynamics, and the corresponding
critical point is described by the $\phi^4$ theory 
in $d+z$ dimensions.
Here, $z$ is the dynamic exponent; $z=3$ for the QCP of the Stoner 
ferromagnet, $z=2$ for the SDW QCP, and $z=1$ for the transverse-field 
Ising model.
Within this framework, an important observation,
first discussed for a QCP of an insulating magnet 
in two dimensions \cite{Chakravarty.89,ChubukovSachdevYe.94},
is that a QCP at $T=0$ will influence physical properties 
at finite temperatures over a finite-range of non-thermal control
parameters. 

While quantum criticality is currently being discussed in a number
of strongly correlated systems \cite{Natphys-qpt08},
it has arguably been most systematically studied in magnetic
heavy fermion metals \cite{Gegenwart.08,HvL-RMP,Coleman-Nature}.
From a materials perspective, the heavy fermions have a number
of advantages in this context.
In particular, the large effective electronic mass -- the defining 
characteristics of these systems -- implies that the relevant 
energy scales are small, making it relatively easy to tune their
ground states by external parameters such as magnetic field 
or pressure \cite{Stewart.00}. 
At the same time,
the understanding of 
their effective Hamiltonian
-- containing competing Kondo and RKKY 
interactions \cite{Hewson,Doniach,Varma76} --
gives us intuition on how the system
is being tuned microscopically 
when an external non-thermal 
control parameter is varied.
Explicit observation of AF QCPs has been made 
in a number of heavy
fermion metals \cite{Gegenwart.08,HvL-RMP},
including, in particular, 
CeCu$_{\rm 6-x}$Au$_{\rm x}$,
YbRh$_{\rm 2}$Si$_{\rm 2}$,
and CePd$_{\rm 2}$Si$_{\rm 2}$.
These systems have 
allowed systematic probes of quantum critical behavior
through transport, thermodynamic, and spectroscopic measurements.
One very basic lesson is that 
the influence of quantum critical fluctuations can cover a large
control-parameter range at non-zero temperatures, and can
extend to surprisingly high 
temperatures \cite{Gegenwart.08,HvL-RMP,Coleman-Nature}.


Theoretically, an important notion that has emerged is that 
QCPs can go beyond the Landau paradigm.
The new type of QCPs being discussed contains critical 
excitations that are inherently quantum-mechanical, in the form
of a critical breakup of Kondo
singlets \cite{Si-Nature,Colemanetal,Senthil.04,PaulPepinNorman.07}.
The notion that there could be emergent quantum excitations beyond 
order-parameter fluctuations is, while un-orthodox, 
in fact natural.
After all, the order parameters we are dealing with 
are coarse-grained classical variables. It is conceivable that 
genuine quantum-mechanical effects -- associated with the Kondo
entanglement effect in our case -- are part of the critical
fluctuations at a QCP. 
These considerations have enjoyed fruitful interactions
with experimental studies in the heavy fermion systems
on spin dynamics \cite{Aronson.95,Schroder,Kadowaki.06,Knafo.09},
Fermi surface \cite{paschen2004,friedemann_hall,shishido2005},
and multiple energy scales \cite{paschen2004,friedemann_hall,gegenwart2007}.
More broadly, they have impacted 
on the 
developments of QCPs in other systems as well,
including in insulating quantum magnets \cite{Senthil-dqcp}.
It appears that the many-body-theory community has largely come 
to terms with the notion that QCPs exist beyond the Landau paradigm.

\section{Kondo lattice and heavy Fermi liquid}
Heavy fermions were traditionally considered as a prototype
for 
a strongly correlated Fermi liquid.
The theory of a heavy Fermi liquid was developed in the 
early 1980s \cite{Hewson}.
The microscopic model for the magnetic heavy fermion materials is 
the Kondo lattice Hamiltonian:
\begin{eqnarray}
{\cal H}
&=&
\frac{1}{2} \sum_{ ij} I_{ij} ~{\bf S}_{i} \cdot {\bf S}_{j} 
~+~\sum_{\bf k \sigma} \epsilon_{\bf k}
c_{{\bf k}\sigma}^{\dagger} c_{{\bf k}\sigma} \nonumber\\
&&+~
\sum_i J_K ~{\bf S}_{i} \cdot {\bf s}_{c,i} 
\;. 
\label{kondo-lattice}
\end{eqnarray}
The model contains a lattice of spin-$1/2$ local moments, 
which interact with each other with an AF 
exchange interaction
$I_{ij}$; we will use $I$ to label say the nearest-neighbor
interaction.
It also includes 
a conduction-electron band, $c_{{\bf k}\sigma}$, with
a band dispersion $\epsilon_{\bf k}$
(and, correspondingly, a hopping matrix $t_{ij}$);
$W$ labels the bandwidth.
At each site $i$, the spin of the conduction
electrons, ${\bf s}_{i,c} = (1/2) c_{i}^{\dagger} \vec{\tau} c_i$,
where $\vec{\tau}$ are the Pauli matrices,
is coupled to a spin-$1/2$ local moment, $\bf {S}_i$,
via an AF Kondo exchange interaction $J_K$.

The Kondo screening effect leads to Kondo resonances,
which are charge-e and spin-$1/2$ excitations.
There is one such Kondo resonance per site, and these excitations 
induce a ``large'' Fermi surface.
Consider that the conduction electron band is filled with $x$ 
electrons per site; for concreteness,
we take $0<x<1$. The conduction electron band and the Kondo resonances
will be hybridized, resulting in a count of $1+x$ electron per site.
The Fermi surface
would therefore
have to expand to a size that encloses 
all these $1+x$ electrons.
This defines the large Fermi 
surface \cite{Hewson,Auerbach-prl86,Millis87,Oshikawa}.

In the regime $I \ll J_K \ll W$, various approaches, in particular
the slave-boson mean field theory \cite{Hewson,Auerbach-prl86,Millis87},
give rise to the following picture.
Consider the conduction electron Green's function:
\begin{eqnarray}
G_c({\bf k},\omega) \equiv F.T.[-<T_{\tau} c_{{\bf k},\sigma}(\tau)
c_{{\bf k},\sigma}^{\dagger}(0)>] \;, 
\label{gc-definition}
\end{eqnarray}
where $F.T.$ is taken with respect to $\tau$. This Green's function
is related to a self-energy, $\Sigma({\bf k},\omega)$, via the standard
Dyson equation:
\begin{eqnarray}
G_c({\bf k},\omega) = \frac{1}{\omega-\epsilon_{\bf k} -
\Sigma({\bf k},\omega)} \;. 
\label{gc-Dyson-equation} 
\end{eqnarray}
In the heavy Fermi liquid state,
$\Sigma({\bf k},\omega)$ is non-analytic and contains
a pole in the energy space:
\begin{eqnarray}
\Sigma({\bf k},\omega)
=\frac{(b^*)^2}{\omega-\epsilon_f^*} \;. 
\label{sigma-pole}
\end{eqnarray}
Inserting Eq.~(\ref{sigma-pole}) into
Eq.~(\ref{gc-Dyson-equation}), we end up with two poles
in the Green's function:
\begin{eqnarray}
G_c({\bf k},\omega) =
\frac{u_{\bf k}^2}{\omega-E_{1,{\bf k}}}
+
\frac{v_{\bf k}^2}{\omega-E_{2,{\bf k}}} \;. 
\label{gc-two-poles}
\end{eqnarray}
Here,
\begin{eqnarray}
E_{1,{\bf k}} &=& (1/2)\left [\epsilon_{\bf k}+\epsilon_f^*
-\sqrt{(\epsilon_{\bf k}-\epsilon_f^*)^2+4(b^*)^2} \right ] \;, 
\nonumber \\
E_{2,{\bf k}} &=&
(1/2) \left [
\epsilon_{\bf k}+\epsilon_f^*
+\sqrt{(\epsilon_{\bf k}-\epsilon_f^*)^2+4(b^*)^2} \; 
\right ]
\label{hf-bands}
\end{eqnarray}
describe the dispersion of the two heavy-fermion bands.
These bands must accommodate $1+x$ electrons, so the new Fermi energy
has to lie in a relatively flat portion of the dispersion, leading
to a small Fermi velocity and a large quasiparticle mass $m^*$.

It is important to make a note here that we have used a
{\it ${\bf k}$-independent self-energy}
to induce a large reconstruction of the quasiparticle dispersion
($\epsilon_{\bf k}$ $\rightarrow$
$E_{1,{\bf k}}, E_{2,{\bf k}}$) and a corresponding large
reconstruction of the Fermi surface. In fact,
the self-energy of Eq.~(\ref{sigma-pole})
contains only two parameters,
the pole strength ({\it i.e.}, the residue), $(b^*)^2$,
and the pole 
location,
$\epsilon_f^*$. 
Eq.~(\ref{sigma-pole}) does not contain the
incoherent features beyond the well-defined pole.
Such incoherent components can be introduced,
through ${\it e.g.}$ the dynamical mean field theory 
\cite{GeorgesRMP,VollhardtKotliar,Jarrell.93,Rozenberg.95},
and they will induce non-zero damping (of the Fermi liquid form)
to the quasiparticle excitations
in Eq.~(\ref{gc-two-poles}).
But the fact remains that a ${\bf k}$-independent self-energy is adequate
to capture the Kondo effect and the resulting heavy quasiparticles. 
We will return to this feature shortly in the discussion of a 
Kondo breakdown effect.

\section{Quantum criticality in the Kondo lattice}
\label{sec:qcp}

\subsection{General considerations}

The Kondo interaction drives the formation of Kondo singlets between
the local moments and conduction electrons. At high temperatures,
the system is in a fully incoherent regime with the local moments
weakly coupled to conduction electrons. Going below some scale
$T_K^0$, the initial screening of the local moments starts to 
set in. Eventually, at temperatures 
below some Fermi-liquid scale, $T_{\rm FL}$, the heavy quasiparticles
are fully developed.

When the AF RKKY interaction among the local moments
becomes larger than 
the Kondo interaction, the system is expected to develop 
an AF order. An AF QCP is then to be expected,
when the control parameter, $\delta = I/T_K^0$, reaches some
critical value $\delta_c$. At $\delta > \delta_c$, the AF order
will develop as the control parameter is lowered through the
AF-ordering line, $T_N(\delta)$.

In addition, the RKKY interactions will also eventually lead 
to the suppression of the Kondo singlets. Qualitatively, 
the RKKY interactions promote singlet formation among 
the local moments, thereby reducing the tendency of singlet
formation between the local moments and conduction electrons.
This will define an energy ($E_{loc}^*$) or temperature ($T_{loc}^*$)
scale, describing the breakdown of the Kondo effect. On very general grounds,
the $T_{loc}^*$ line is expected to be a crossover at non-zero temperatures,
but a sharp transition at zero temperature.

To study these issues theoretically, one key question is how 
to capture not only the magnetic order and Kondo-screening, but also 
the dynamical competition between the Kondo and RKKY interactions.
The microscopic approach
that is capable of doing this is the extended dynamical mean
field theory (EDMFT) \cite{SiSmith.96,SmithSi.00,Chitra.00}.
The two solutions \cite{Si-Nature,Si-prb.03,Grempel.03,ZhuGrempelSi.03,SunKotliar.03,Glossop07,Zhu07,Glossop_etal09}
that have been derived through EDMFT are
summarized below. 

Large-N approaches based on slave-particle representations of the spin
operator are also commonly used to study Kondo-like systems. 
One type of approach is based on a fermionic representation 
of the spin \cite{Senthil.04,PaulPepinNorman.07}. This representation
naturally incorporates the physics of singlet formation, so it 
captures the Kondo singlets (as well as the singlets among the local
moments), but it does not allow magnetism in the
large-N limit. One may allow a magnetic order in a static mean-field
theory for a finite-N \cite{Senthil.04}. However, 
the magnetic transition and breakdown of Kondo screening are 
always separated in the phase diagram and the zero-temperature 
magnetic transition is still of the SDW type. This, we believe, 
is a manifestation of the static nature of the mean field theory.

A Schwinger-boson-based large-N formulation is another microscopic
approach that is being considered in this context \cite{Rech}. 
This approach naturally incorporates 
magnetism. While it is traditionally believed that bosonic
representations of spin in general have difficulty to capture 
the Kondo screening physics at its large-N limit, there is 
indication \cite{Rech} that the dynamical nature of the formulation
here allows an access to at least aspects of the Kondo effect.
It will be interesting to see what type of quantum phase transitions
this approach will lead to for the Kondo lattice problem.

This is a subject that is still in a very rapid development, 
and a number of other theoretical approaches are also being 
taken \cite{Pepin.07,Deleo.08,Nevidomskyy.09,Paul.09}.

\subsection{Microscopic approach based on the extended dynamical mean field theory}

The EDMFT method incorporates intersite collective fluctuations
into the dynamical mean field theory framework 
\cite{GeorgesRMP,VollhardtKotliar}.
This systematic method is constructed within
a ``cavity'', diagrammatic, or functional 
formalism~\cite{SiSmith.96,SmithSi.00,Chitra.00}.
It is conserving, satisfying the various Ward identities.
Diagrammatically, EDMFT 
incorporates an infinite series
associated with intersite interactions,
in addition to
the local processes already taken into account in the 
dynamical mean field theory.


Within the EDMFT, the dynamical spin susceptibility
and the conduction-electron Green's function
respectively have the forms:
$\chi ({\bf q}, \omega) = \frac {1}  
{ M(\omega) + I_{{\bf q}}} $,
and 
$G ({\bf k}, \epsilon) = \frac {1}  {\epsilon + \mu
- \epsilon_{\bf k} - \Sigma (\epsilon)} $ .
The correlation functions, 
$\chi ({\bf q}, \omega)$ and 
$G ({\bf k}, \epsilon)$, are momentum-dependent.
At the same time, the irreducible quantities,
$ M(\omega)$ and 
$\Sigma (\epsilon) $ are momentum-independent.
They are determined in terms of 
a 
Bose-Fermi Kondo model,
\begin{eqnarray}
{\cal H}_{\text{imp}} &=& J_K ~{\bf S} \cdot {\bf s}_c +
\sum_{p,\sigma} E_{p}~c_{p\sigma}^{\dagger}~ c_{p\sigma}
\nonumber \\
&& + \; g \sum_{p} {\bf S} \cdot \left( {\bf \Phi}_{p} +
{\bf \Phi}_{-p}^{\;\dagger} \right) + 
\sum_{p}
w_{p}\,{\bf \Phi}_{p}^{\;\dagger} \cdot {\bf \Phi}_{p}\;. \nonumber
\\ \label{H-imp}
\end{eqnarray}
The fermionic and bosonic baths are determined by self-consistency
conditions, which manifest the translational invariance,
$\chi_{{loc}} (\omega) 
= \sum_{\bf q} \chi ( {\bf q},
\omega )$,
and $G_{{loc}} (\omega) = \sum_{\bf k} G( {\bf k}, \omega )$.

The $0+1$-dimensional quantum impurity problem,
Eq.~(\ref{H-imp}), has the following Dyson equations:
$M(\omega)=\chi_{0}^{-1}(\omega) + 1/\chi_{\rm loc}(\omega)$
and $\Sigma(\omega)=G_0^{-1}(\omega) - 1/G_{\rm loc}(\omega)$, where
$\chi_{0}^{-1} (\omega) = -g^2 \sum_p 2 w_{p} /[\omega^2 -
w_{p}^2]$ and $G_0 (\omega) = \sum_p 1/(\omega - E_p)$ are the
Weiss fields.
The EDMFT formulation allows us to study different degrees of quantum
fluctuations as manifested in the spatial dimensionality of these
fluctuations. The case of two-dimensional 
magnetic fluctuations are represented in terms of
the RKKY density of states that has a non-zero value at the lower edge,
{\it eg.}:
\begin{eqnarray}
\rho_{I} (\epsilon) \equiv  \sum_{\bf q} \delta ( \epsilon  - I_{\bf q} )
= (1/{2 I}) \Theta(I - | \epsilon | ) \;,
\label{rho_I_2D}
\end{eqnarray}
where $\Theta$ is the Heaviside step function.
Likewise, three-dimensional magnetic fluctuations are described 
in terms of $\rho_{I} (\epsilon)$ which vanishes at the lower edge
in a square-root fashion, for example:
\begin{eqnarray}
\rho_{I} (\epsilon) = (2/{\pi I^2}) \sqrt{I^2-\epsilon^2}
\Theta(I - | \epsilon | ) \;.
\label{rho_I_3D}
\end{eqnarray}

The bosonic bath reflects the effect of the dynamical magnetic 
correlations, primarily among the local moments, on the local Kondo
effect. On approach to a magnetic quantum critical point, its 
spectrum turns soft, and its ability to suppress the Kondo effect
increases. This effect has been explicitly seen in a number of specific 
studies \cite{Si-Nature,Si-prb.03,Grempel.03,ZhuGrempelSi.03,SunKotliar.03,Glossop07,Zhu07,Glossop_etal09}.
Moreover, the zero-temperature transition 
is second-order whenever the same form of the 
effective RKKY interaction appears in the formalism on both
sides of the transition \cite{SiZhuGrempel.05,SunKotliar.05}.

%
%

\subsection{Spin-density-wave QCP}


The reduction of the Kondo-singlet amplitude 
by the dynamical effects of the RKKY interactions among the local
moments has been considered in some detail in a number of 
studies based on EDMFT 
\cite{Si-Nature,Si-prb.03,Grempel.03,ZhuGrempelSi.03,SunKotliar.03,Glossop07,Zhu07,Glossop_etal09}.
Irrespective of the spatial dimensionality, this weakening of the 
Kondo effect is seen through the reduction of an $E_{loc}^*$ scale.

Two classes of solutions emerge depending on whether this
Kondo breakdown scale vanishes at the AF QCP. 
In the case of Eq.~(\ref{rho_I_3D}), $E_{loc}^*$ has not yet been 
completely suppressed to zero when the AF QCP,
$\delta_c$, is reached from the paramagnetic 
side (but it can go to zero inside the AF region, as further discussed
in Sec.~\ref{sec:af_kl}).
The quantum critical behavior, at energies below 
$E_{loc}^*$, falls within the Hertz-Moriya-Millis type
\cite{Hertz,Moriya,Millis}.
The zero-temperature dynamical spin susceptibility has the 
following form:
\begin{eqnarray}
\chi({\bf q}, \omega ) =
\frac{1}{f({\bf q}) - ia \omega}
\;.
\label{chi-qw-sdw}
\end{eqnarray}
Here $f({\bf q})=I_{\bf q}-I_{\bf Q}$, and is generically $\propto 
({\bf q}-{\bf Q})^2 $ as the wavevector ${\bf q}$ approaches 
the AF ordering wavevector ${\bf Q}$.
The QCP is described by a Gaussian fixed point. At non-zero temperatures,
a dangerously irrelevant operator invalidates the so-called $\omega/T$
scaling \cite{Millis}.

\subsection{Local quantum critical point}

Another class of solution corresponds to $E_{loc}^*=0$ already at $\delta_c$.
It arises in the case of Eq.~(\ref{rho_I_2D}), where the quantum critical
magnetic fluctuations are strong enough to suppress the Kondo effect.


\begin{figure}[t]%
\includegraphics*[width=\linewidth,height=\linewidth]{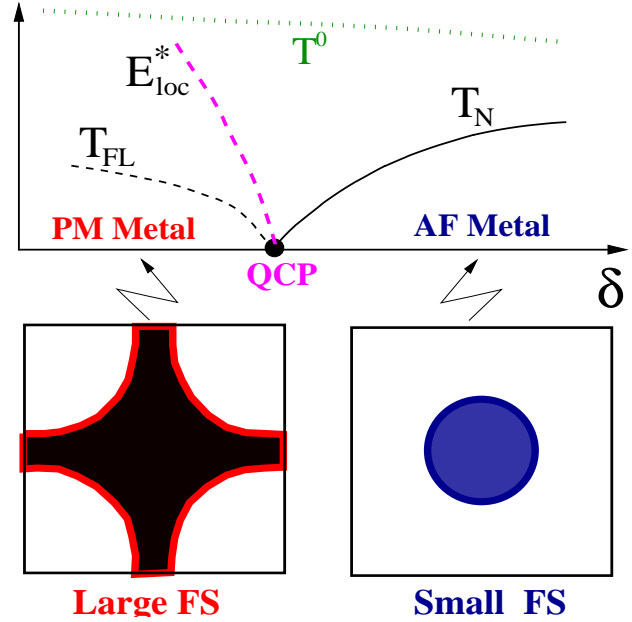}%
\caption{%
Illustration of the local quantum critical point,
which has a critical breakdown of the Kondo effect.
The $E_{loc}^*$ scale separates between the regimes where 
the system goes towards either the Kondo-screened paramagnetic
metal ground state or the Kondo-destroyed
AF metal ground state.
The Fermi surface goes from being large in the paramagentic
metal phase to being small in the AF metal phase.
$T^0$ is the scale at which the initial Kondo screening sets in as 
temperature is lowered from above $T^0$.
}
    \label{lqcp}
\end{figure}

The solution to the local spin susceptibility has the form:
\begin{eqnarray}
\chi({\bf q}, \omega ) =
\frac{1}{f({\bf q}) + A \,(-i\omega)^{\alpha} W(\omega/T)}\;.
\label{chi-qw-T}
\end{eqnarray}
This expression was derived \cite{Si-Nature,Si-prb.03}
within the EDMFT studies,
and through the aid of an $\epsilon$-expansion approach 
to the Bose-Fermi Kondo model.
At the AF QCP,
the Kondo effect itself is critically destroyed ({\it cf.} 
Fig.~\ref{lqcp}).
The calculation of the critical exponent $\alpha$ is
beyond the reach of the $\epsilon$-expansion. In the
Ising-anisotropic case, 
$\alpha$ has 
been 
calculated numerically 
\cite{Grempel.03,Glossop07,Zhu07,Glossop_etal09}:
it is fractional,
and is about $0.7$. 

The breakdown of the Kondo effect not only affects magnetic
dynamics, but also influences the single-electron excitations.
As the QCP is approached from the paramagnetic side, the quasiparticle
residue $z_L \propto b^*$, the strength of the pole
of $\Sigma({\bf k},\omega)$ [{\it cf.} Eq.~(\ref{sigma-pole})], 
goes to zero. The large Fermi surface turns critical.

The breakdown of the large Fermi surface suggests that the Fermi
surface will be small on the antiferromagnetically ordered side.
This leads us to the issue of the Kondo effect inside the AF phase, 
which we now discuss in some detail.

%

\section{Antiferromagnetism and Fermi surface in Kondo lattices}
\label{sec:af_kl}



To consider the Kondo effect in the ordered phase,
we focus on the parameter regime of the Kondo lattice model,
Eq.~(\ref{kondo-lattice}), in the 
the limit $J_K \ll I \ll W$. Here $I$ is the scale for 
the RKKY exchange interaction, and $W$ the bandwidth of the
conduction-electron band.

In this limit, we can use as our reference point the $J_K=0$ 
case \cite{Yamamoto07}.
At this reference point, the local moments with AF exchange interactions
are decoupled from the conduction electrons. 
We will focus on the case that the local-moment
system itself is in a collinear AF state.
The low-energy theory for the local-moment Hamiltonian 
[the first term of Eq.~(\ref{kondo-lattice})]
is the quantum non-linear sigma model (QNL$\sigma$M) 
\cite{Haldane.83,Chakravarty.89}:
\begin{eqnarray}
{\cal S}_{\text{QNL}\sigma\text{M}}
= \frac{c}{2g}\int d^dxd\tau\left[ \left(\nabla 
{\bf n}
\right)^2 +
\left(\frac{\partial
{\bf n}}{ c ~\partial\tau}\right)^2 \right]
\; .
\label{qnlsm}
\end{eqnarray}
Here $c$ is the spin-wave velocity, and $g$ describes the quantum 
fluctuations. 
There are gapless excitations in two regions of the wavevector
space: the staggered magnetization
(${\bf q}$ near ${\bf Q}$) specified by the ${\bf n}$
field and the uniform magnetization (${\bf q}$ near ${\bf 0}$)
described by ${\bf n} \times \partial {\bf n}/\partial \tau$. 

When the Fermi surface of the
conduction electrons does not intersect the AF zone boundary,
only the uniform component of the local moments 
can be coupled to the 
spins of the conduction-electron states near the Fermi surface.
The effective Kondo coupling takes the form,
\begin{eqnarray}
{\cal S}_K = 
\lambda\int d^dx d\tau
~\vec{s}_c
\cdot 
{\bf n} \times \partial {\bf n}/\partial \tau
\label{kondo-lambda}
\end{eqnarray}

A momentum-shell RG treatment requires a procedure 
that mixes bosons, which scale along all directions in momentum
space \cite{Wilson.74}, and fermions, which scale along the
radial direction perpendicular to the Fermi surface \cite{shankar1994}.
Using the procedure specified in Ref.~\cite{Yamamoto09},
we found $\lambda$ to be marginal at the leading 
order 
\cite{Yamamoto07}, just like in the paramagnetic case.
The difference from the latter appears at the loop level:
$\lambda$ is exactly marginal to infinite loops \cite{Yamamoto07}.

The fact that $\lambda$ does not run towards infinity implies
a breakdown of the Kondo effect. This is supplemented 
by a large N calculation \cite{Yamamoto07}, which showed
that the effective
Kondo coupling, Eq.~(\ref{kondo-lambda}), leads to the following
self-energy for the conduction electrons:
\begin{eqnarray}
\Sigma({\bf k},\omega)
\propto \omega^d \; .
\label{sigma-no-pole}
\end{eqnarray}
The absence of a pole in $\Sigma({\bf k},\omega)$, in contrast to
Eq.~(\ref{sigma-pole}), implies the absence of any Kondo
resonance. Correspondingly, the Fermi surface is small.

When the Fermi surface of the conduction electrons intersects the 
AF zone boundary \cite{Yamamoto.08}, 
the staggered magnetization, ${\bf n}$, can be directly
coupled to the spins of the conduction electrons. However,
this coupling is proportional to
${\bf q-Q}$,
as dictated by Adler's theorem.
The Kondo coupling remains marginal, and the Fermi surface
remains small.

\section{Towards a global phase diagram}
\label{sec:global}

\subsection{How to melt a Kondo-destroyed antiferromagnet}

Given the understanding that the AF state with a small Fermi surface
is a stable phase, it would be illuminating to approach the 
quantum transition from this ordered state.

One may consider to use the QNL$\sigma$M representation, and study
the transition by increasing the effective Kondo coupling, $\lambda$
of Eq.~(\ref{kondo-lambda}). 
A recent study, using an RG analysis of an action closely
related to that discussed in Sec.~\ref{sec:af_kl} but with the
conduction electrons
integrated out, has gone along this direction \cite{Ong_Jones09}.
Such an analysis,
however, cannot capture the overall phase diagram of the Kondo lattice
systems. What is missing so far is 
the Berry-phase
term of the QNL$\sigma$M representation:
\begin{eqnarray}
{\cal S}_{\text{Berry}} 
&=& i~s~\sum_{\bf x} \eta_{\bf x} A_{\bf x} 
\nonumber\\
A_{\bf x} &=&
\int_0^{\beta}d\tau
\int_0^1 du 
\left
[ 
{\bf n}\cdot
\left(\frac{\partial {\bf n}}{\partial u}
\times 
\frac{\partial 
{\bf n}}
{\partial \tau} \right) \right]
\label{berry-phase}
\end{eqnarray}
Here, $s=\frac{1}{2}$ is the size of the local-moment 
spin,  $A_x$ is the area on the unit sphere spanned by 
${\bf n}({\bf x},\tau)$ with $\tau \in (0,\beta)$,
and $\eta_x=\pm 1$ at even/odd sites (consider, for definiteness, 
a cubic or square lattice with N\'{e}el order).

The Berry phase term can be neglected deep inside the AF phase.
For smooth configurations of ${\bf n}$ in the $({\bf x},\tau)$
space, the Berry phase term vanishes. Topologically non-trivial
configurations of ${\bf n}$ in $({\bf x},\tau)$ yield a finite Berry
phase. They, however,
cost a non-zero energy inside the AF phase and can  
be neglected 
for small $J_K$ and, correspondingly, small $\lambda$.
As $J_K$ is increased, 
on the other hand,
these gapped configurations 
come into play. Indeed, they are expected to be crucial for
capturing the Kondo effect.
Certainly, the Kondo singlet formation requires the knowledge of the size
of the microscopic spins, and the Berry phase term is what encodes
the size of the spin in the QNL$\sigma$M representation. 

\subsection{Global phase diagram}

\begin{figure}[t]%
\includegraphics*[width=\linewidth,height=0.9\linewidth]{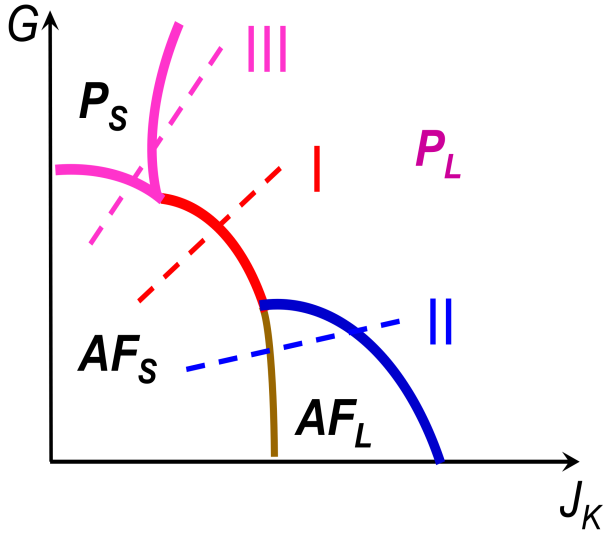}%
\caption{%
The $T=0$ global phase diagram of the AF
Kondo lattice. $G$ describes magnetic frustration or 
spatial dimensionality, and $j_K$ is the normalized 
Kondo coupling. The four ($2^2$) types of phases,
${\rm AF_S}$, ${\rm AF_L}$, ${\rm P_S}$, and ${\rm P_L}$ arise since 
they contain two kinds of distinctions:
antiferromagnetism (AF) or paramagnetism (P) on the one hand,
and 
Kondo screening (Fermi surface large, ``L'') 
or Kondo breakdown (Fermi surface small, ``S'') on the other hand.
The lines ``$I$'', ``$II$'', ``$III$'' describe three types of
trajectories for the quantum transition.
}
    \label{global_pd}
\end{figure}

We can address these effects at a qualitative level, in terms of a 
global phase diagram. We consider a two-dimensional 
parameter space \cite{Si-physicab-06}, as shown in Fig.~\ref{global_pd}.
The vertical axis describes the local-moment magnetism. It 
is parametrized by $G$,
which characterizes the degree of quantum fluctuations 
of the local-moment magnetism;
increasing $G$ reduces the N\'{e}el order.
This parameter can be a measure of magnetic frustration,
{\it e.g.} $G=I_{\rm nnn}/I_{\rm nn}$, the ratio of the
next-nearest-neighbor exchange interaction to the nearest-neighbor
one, or it can be the degree of spatial anisotropy.
The horizontal axis is $j_K\equiv J_K/W$, the Kondo 
coupling normalized by the conduction-electron bandwidth.
We are considering a fixed value of $I/W$, which is typically
much less than $1$, and a fixed number of conduction electrons 
per site, which is taken to be $0<x<1$ 
without
a loss of generality.
 
The ${\rm AF_S}$ phase describes the small-Fermi-surface 
AF state, whose existence has been established
asymptotically exactly using the RG method as described in the
previous section.
The ${\rm P_L}$ phase is the standard heavy Fermi liquid 
with heavy quasiparticles and a large Fermi surface \cite{Hewson}.
The ${\rm AF_L}$ phase corresponds to an AF state 
in the presence of Kondo screening. It can either be considered 
as resulting from the ${\rm AF_S}$ phase once the Kondo screening sets in,
or from the ${\rm P_L}$ phase via an SDW instability.

We have discussed the above three phases in some detail 
before \cite{Si-physicab-06,Yamamoto07}. Also alluded to
in Ref.~\cite{Si-physicab-06}
is another important feature of this global phase diagram.
Along the vertical axis, the conventional N\'{e}el order becomes
unstable as $G$ goes beyond some threshold value.
The phase at $G>G_c$ is a paramagnet, restoring the 
spin-rotational invariance that was broken in the 
N\'{e}el state. It typically still contains valence-bond solid order
(spin Peierls),
although the cases without any conventional-symmetry-breaking (spin
liquid) have also been extensively discussed 
especially in lattices
with strong geometrical frustration.

These considerations lead to the 
natural possibility of a ${\rm P_S}$ phase, a paramagnetic
phase with a Kondo breakdown 
(and, hence, a small Fermi surface) which
either breaks or preserves translational invariance.
Related Considerations are also being pursued in
Refs.~\cite{Custers.09,Coleman.09}.

This global phase diagram contains three routes for a system to 
go from the ${\rm AF_S}$ phase to the ${\rm P_L}$ phase.

\begin{itemize}

\item Trajectory ``I'' is a direct transition between the two.
This ${\rm AF_S}-{\rm P_L}$ transition gives rise to a local quantum 
critical point. A critical Kondo breakdown occurs at the 
AF QCP, giving rise to 
a sudden small-to-large jump of the Fermi surface 
and the vanishing of a Kondo-breakdown scale,
$E_{loc}^*$ \cite{Si-Nature,Colemanetal}.

\item Trajectory ``II'' goes through the ${\rm AF_L}$ phase. The quantum
critical point at the ${\rm AF_L}-{\rm P_L}$ 
boundary falls in the Hertz-Moriya-Millis
type \cite{Hertz,Moriya,Millis}.

\item Trajectory ``III'' goes through the ${\rm P_S}$ phase. 
The ${\rm P_S}-{\rm P_L}$
transition could describe either a spin liquid to heavy Fermi liquid 
QCP \cite{Senthil.04,PaulPepinNorman.07},
or a spin Peierls to heavy Fermi liquid QCP.

\end{itemize}

\subsection{Discussion of the global phase diagram}
\label{sec:connection}

We now turn to a number of points to 
elaborate on the global phase diagram.

Consider first the transition along the trajectory ``I''. 
For this transition to be second order, the quasiparticle 
residues associated with both the small and large Fermi 
surfaces must vanish as the QCP is approached
from either side \cite{Si-Nature,Colemanetal,Si-prb.03,Si-physicab-06,Senthil.06}.

The transition along the trajectory ``II'' involves an intermediate
${\rm AF_L}$ phase. This transition can in general be 
specified \cite{Yamamoto09b} through an ``order parameter''
for the Kondo screening,
$<\sum_{\sigma}F_{\sigma}^{\dagger}c_{\sigma}> \ne 0$
(where $F_{\sigma}$ is a composite operator involving 
a spin operator of the local moment and a conduction electron
operator),
which is non-zero in the ${\rm AF_L}$ phase and vanishes in the 
${\rm AF_S}$. When the AF order parameter 
is relatively small, this transition coincides with a 
Lifshitz transition between Fermi surfaces 
of different topology \cite{Si-physicab-06,Yamamoto07}. 
Several authors \cite{Watanabe07,Martin08,Vojta08,Lanata08}
also identified a Lifshitz transition within a static
slave-boson mean-field and related treatments \cite{Lacroix.79,ZhangYu.00}
of the Kondo lattice
problem. In our global phase diagram, these transitions should
be considered as transitions inside the ${\rm AF_L}$ region
as the AF order parameter is increased \cite{Yamamoto09b}, and are
to be differentiated from the ${\rm AF_L}$ to ${\rm AF_S}$ transition discussed
here.

For a continuous transition between the ${\rm AF_S}$ and 
${\rm AF_L}$ phases,
the quasiparticle residues of both the large and small Fermi surfaces
must again vanish continuously as the QCP is approached 
from either side. 

The ${\rm P_S}$ phase is a state with suppressed Kondo effect.
Whether it corresponds to a non-Fermi liquid phase \cite{Anderson.08}
or displays Fermi liquid behavior is an issue that remains to be
determined. Two factors come into play. First, what is the nature
of the local-moment component? In a two-dimensional square lattice,
for instance, spin-$1/2$ moments
at $G>G_c$ are expected to develop a spin-Peierls order. With geometrical
frustration, as occurring in {\it eg.} a pyrochlore or kagome lattice, 
it is possible that the local-moment component goes into a spin liquid phase,
although this issue is not yet theoretically settled. 
Second, what is the nature of the Kondo coupling between the local moments
and the conduction electrons? The answer will crucially depend on 
whether the excitation spectrum of the local-moment component 
is gapped or gapless. If it is gapped, the Kondo coupling will be irrelevant
and the low-energy properties will have a Fermi liquid form. 
If it is gapless, the behavior of the Kondo coupling can vary
depending on its 
form 
expressed in terms of the 
low-energy effective excitations of the local-moment component
and the conduction electrons. 

\section{Experiments on AF heavy fermions}
\label{sec:expt-AF}

There are by now many heavy fermion metals in which QCPs have been
either explicitly observed, or implicated. We discuss some of them
in light of our theoretical considerations.

\subsection{Global phase diagram}

A number of heavy fermion materials 
might be classified according to our global phase diagram, Fig.~\ref{global_pd}.

In CeCu$_{\rm 6-x}$Au$_{\rm x}$, both the pressure- and 
doping-induced QCPs show the characteristics 
of local quantum criticality, accessed through trajectory ``I''.
The field-induced QCP \cite{Stockert_CeCuAu_field},
however, has the properties of an 
SDW QCP. We interpret the field-tuning as taking the trajectory
``II''. It will be interesting to explore 
whether an ${\rm AF_S}$-${\rm AF_L}$
boundary can be located as a function of magnetic field.

Perhaps the most complete information exists in the pure and 
doped YbRh$_{\rm 2}$Si$_{\rm 2}$
system. 
In the pure YbRh$_{\rm 2}$Si$_{\rm 2}$,
strong evidence exists that
the field-induced transition goes along the 
trajectory ``I'' (see below). A surprising recent development came
from experiments in the doped YbRh$_{\rm 2}$Si$_{\rm 2}$.
In the Co-doped YbRh$_{\rm 2}$Si$_{\rm 2}$, the field-induced
transition seems to travel along trajectory ``II'' \cite{Friedemann09}.
In the Ir-doped \cite{Friedemann09} and Ge-doped \cite{Custers.09} 
YbRh$_{\rm 2}$Si$_{\rm 2}$, 
on the other hand, the field-induced transition appears
to go along trajectory ``III''.
Recent experiments in pure YbRh$_{\rm 2}$Si$_{\rm 2}$
under pressure \cite{Tokiwa09} yield results which are very
similar to those of Co-doped YbRh$_{\rm 2}$Si$_{\rm 2}$
at ambient pressure, suggesting that the results observed in
the doped YbRh$_{\rm 2}$Si$_{\rm 2}$ are in fact intrinsic
and do not primarily result from disorder.

CeIn$_3$ is one of the earliest heavy fermion metals in which
an AF QCP was implicated \cite{Mathur98}. This system is cubic,
and we would expect it to lie in the small $G$ region of the 
global phase diagram. Indeed, there is indication
that this cubic material displays an ${\rm AF_S}$-${\rm AF_L}$ Lifshitz
transition as a function of magnetic field \cite{Sebastian.09}.

It is to be expected that magnetic frustration will help reach 
the ${\rm P_S}$ phase.
The heavy fermion system 
${\rm YbAgGe}$ has a hexagonal lattice,
and, indeed, there is some indication 
that the ${\rm P_S}$ phase exists in this
system~\cite{Canfield}; lower-temperature measurements over an extended 
field range, however, will be needed to help establish the
detailed phase diagram.

Finally, it is important to note that considerable experiments 
exist on the nature of the Fermi surface inside the various 
phases. The ${\rm P_L}$ phase was historically 
established through the observation of the large heavy-fermion
Fermi surface \cite{TailleferLonzarich.88},
while the existence 
of the ${\rm AF_S}$ phase itself has been supported by the 
Fermi-surface measurements
in a large number of AF heavy fermions \cite{McCollam.05,Onuki.03}.

\subsection{Kondo breakdown at the antiferromagnetic QCP}

The most direct evidence for the local quantum critical point 
occurs in YbRh${\rm _2}$Si${\rm _2}$ and in CeCu$_{\rm 6-x}$Au$_{\rm x}$.
For YbRh${\rm _2}$Si${\rm _2}$, the Fermi-liquid behavior is observed 
both inside the AF-ordered phase and the field-induced non-magnetic
phase \cite{Custers.03}. In addition, Hall-coefficient measurements
\cite{paschen2004,friedemann_hall} have provided fairly direct evidence 
for the breakdown of the Kondo effect precisely at the AF QCP.
The existence of the Kondo-breakdown scale, $T_{loc}^*$, has 
also been seen in both the Hall \cite{paschen2004,friedemann_hall} 
and thermodynamic \cite{gegenwart2007} experiments.

For CeCu$_{\rm 6-x}$Au$_{\rm x}$, the unusual magnetic 
dynamics \cite{Schroder}
observed near the $x =x_c \approx 0.1$
by early
neutron scattering measurements 
is understood in terms of such a critical Kondo breakdown
in the form of local quantum criticality.
A divergent effective mass expected in this picture is consistent
with the thermodynamic measurement in both the doping and 
pressure induced QCP in this system \cite{HvL-RMP}.
This picture necessarily implies a Fermi-surface jump across
the QCP, as well as a Kondo-breakdown energy scale 
$E_{loc}^*$ going to zero at the QCP, but such characteristics are
yet to be probed in CeCu$_{\rm 6-x}$Au$_{\rm x}$.

CeRhIn$_{\rm 5}$ is a member of the Ce-115 heavy fermions \cite{Hegger.00}.
It contains both antiferromagnetism and superconductivity 
in its pressure-field phase diagram. 
When a large-enough magnetic-field is applied and superconductivity
is removed ($H>H_{c2}$), there is evidence for a single 
QCP between antiferromagnetic to non-magnetic 
phases \cite{park-nature06,Knebel.08}.
At this QCP, the de Haas-van Alphen (dHvA) results \cite{shishido2005}
suggest a jump
in the Fermi surface and a divergence in the effective mass.

One of the earliest systems in which anomalous magnetic dynamics 
was observed is ${\rm UCu_{5-x}Pd_x}$ \cite{Aronson.95}.
It is tempting to speculate \cite{Si-physicab-06} that a Kondo-destroying 
spin-glass QCP underlies this observation.

\subsection{Spin-density-wave QCP}

Neutron scattering experiments have provided some evidence that 
the AF QCPs in 
both 
Ce(Ru${\rm _{1-x}}$Rh$_{\rm x}$)$_{\rm 2}$Si${\rm _2}$ \cite{Kadowaki.06}
and Ce${\rm _{1-x}}$La$_{\rm x}$Ru$_{\rm 2}$Si${\rm _2}$ \cite{Knafo.09}
have the SDW form.
Likewise,
in CeCu$_{\rm 2}$Si${\rm _2}$,
transport and thermodynamic measurements \cite{Gegenwart.98}
have indicated that
its field-induced QCP 
belongs to the SDW category.

\section{Ferromagnetic phases and phase transitions}
\label{sec:ferro}

Compared to their AF counterpart, quantum phase transitions in 
ferromagnetic heavy fermions have hardly received theoretical 
attention. While ferromagnetic heavy fermions are rarer to 
begin with, the list is steadily growing and there are by
now more than a dozen such systems being studied 
experimentally.
While 
the same physics believed to make
ferromagnetic quantum transitions first order 
in the case of 
weak magnets \cite{belitz-rmp05,Abanov_Chubukov04}
may also prevail here, it is still important to address whether the 
Kondo-breakdown physics also appears in the ferromagnetic heavy fermions.

In a recent study \cite{yamamoto_fm_kl}, 
we have considered the Kondo lattice model
in which the direct exchange between the local moments is
ferromagnetic while the Kondo interaction is still antiferromagnetic.
It turns out that, due to a separation of energy scales, the ferromagnetic
order in the parameter regime, $J_K \ll |I| \ll W$, 
is amenable to an RG analysis. 
A ferromagnetically ordered phase with a Kondo-breakdown small Fermi
surface is seen to be stable, even though the conduction electrons
and local moments are strongly coupled to each other.
In this ${\rm F_S}$ phase, non-Fermi liquid behavior 
appears over an appreciable
range of frequencies and temperatures. These results provide the basis
to understand some long-standing puzzles associated with the 
dHvA observation of a small Fermi surface
in some ferromagnetic heavy fermion metals \cite{King.91}.
They may also be related to the non-Fermi liquid behavior
observed in the ferromagnetically-ordered state of
URu$_{2-x}$Re$_x$Si$_2$ \cite{Bauer05}.
Finally, they 
raise the prospect for a Kondo-breakdown-type 
ferromagnetic quantum phase transition.

\section{Summary}

We have discussed the physics beyond the order-parameter fluctuations
in the quantum criticality of Kondo lattice systems. Of particular
interest is the local quantum critical point, which features a critical
Kondo breakdown. Microscopic studies on this type of quantum critical
point have mostly been based on the extended dynamical mean field theory.

The critical Kondo breakdown at the local quantum critical point leads
to a jump in the Fermi surface, a critical suppression of the quasiparticle
residues of both the small and large Fermi surfaces, and the vanishing 
of a Kondo-breakdown scale at the quantum critical point. Considerable 
experimental evidences for such properties have emerged, which we have 
summarized.

At the local quantum critical point, there is also a dynamical spin
susceptibility which has features of an interacting
fixed point. The form, given in Eq.~(\ref{chi-qw-T}),
contains a self-energy that has an anomalous frequency dependence,
with a fractional exponent, as well
as $\omega/T$ scaling. In contrast to the non-analytic frequency dependence,
the momentum dependence of the self-energy is completely regular [which is,
in fact, absent in Eq.~(\ref{chi-qw-T})].
It is interesting to note that recent studies of quantum critical behavior 
using the gravitational perspective developed in the string-theory context 
have identified certain symmetry reasons \cite{Faulkner.09} for 
a single-electron self-energy
with non-analytic frequency
dependence and smooth momentum dependence \cite{Varma89}.
Whether related emergent symmetry can be identified within
the gravitational description
for the contrasting frequency and momentum dependence
in the two-particle self-energy,
as appearing in 
Eq.~(\ref{chi-qw-T}),
is an intriguing open 
question worthy of future studies.

We have also discussed a global phase diagram for the magnetic 
heavy fermion metals. Detailed theoretical studies to access 
the overall phase diagram will be much needed. 

These developments on the new type 
of magnetic quantum phase 
transitions with unusual evolutions of the Fermi 
surface have not yet been accompanied 
by corresponding studies on superconductivity.
There are many theoretical questions one can ask.
Microscopically,
whether and how superconductivity can arise near
the Kondo-breakdown local quantum critical 
point \cite{Gegenwart.08,park-nature06} is an intriguing open question.
Macroscopically, there are general 
considerations 
that the entropy accumulation near quantum critical 
points \cite{ZhuGarst.03,Rost.09} foster the formation of 
unconventional phases, including 
unconventional superconductivity \cite{Mathur98}. It would be quite
meaningful to explore whether a theoretical framework can be 
developed to implement such considerations.

\begin{acknowledgement}
I am grateful to E. Abrahams, R. Bulla, 
J. Dai, M. Glossop, P. Goswami, D. Grempel, 
K. Ingersent,
S. Kirchner, E. Pivovarov, J. Pixley, 
S. Rabello, J. L. Smith, J. Wu, S. Yamamoto, 
J.-X. Zhu, and L. Zhu for collaborations on the
theoretical aspects of this subject.
I would also like to thank many colleagues 
for discussions, particularly P. Coleman and 
F. Steglich on the global phase diagram.
This work has been supported in part by the NSF 
Grant No. DMR-0706625 and the Robert A. Welch 
Foundation Grant No. C-1411. 
\end{acknowledgement}

%
 \bibliographystyle{pss}
 \bibliography{qcp_gpd}


\end{document}